\pdfoutput=1 
\documentclass[12pt]{article}
\usepackage{eprintsty}

\newcommand\pubdate{\today}
\newcommand\pubnumber{CP3-21-67}
\newcommand{\OO}{\ensuremath{\mathcal{O}}}

\newcommand{\sss}{\scriptscriptstyle}
\newcommand{\Op}[1]{\OO_{\sss #1}}
\newcommand{\Opp}[2]{\OO_{\sss #1}^{\sss #2}}

\renewcommand{\phi}{\ensuremath{\varphi}}

\def\Title#1{\begin{center} {\Large #1 } \end{center}}
\def\Author#1{\begin{center}{ \sc #1} \end{center}}
\def\Address#1{\begin{center}{ \it #1} \end{center}}

\newcommand\pubblock{\rightline{\begin{tabular}{l} \pubnumber\\
         \pubdate  \end{tabular}}}
\newenvironment{Abstract}{\begin{quotation}  }{\end{quotation}}
\newenvironment{Presented}{\begin{quotation} \begin{center} 
             PRESENTED AT\end{center}\bigskip 
      \begin{center}\begin{large}}{\end{large}\end{center} \end{quotation}}
\def\Acknowledgements{\bigskip  \bigskip \begin{center} \begin{large}
             \bf ACKNOWLEDGEMENTS \end{large}\end{center}}

\begin{document}
\begin{titlepage}
\pubblock

\vfill
\Title{$tWZ$ production at NLO in QCD in the SMEFT}
\vfill
\Author{Hesham El Faham}
\Address{Centre for Cosmology, Particle Physics and Phenomenology (CP3), Universit\'e catholique de Louvain, B-1348 Louvain-la-Neuve, Belgium and Vrije Universiteit Brussel, Pleinlaan 2, 1050 Brussels, Belgium}
\vfill
\begin{Abstract}
At NLO in QCD, the $tWZ$ production process interferes with resonant LO processes, $t\bar{t}Z$ and $t\bar{t}$, and a method to meaningfully disentangle these overlapping processes needs to be employed. I discuss the diagram-removal procedure needed to obtain an operative definition of $tWZ$ at NLO accuracy. I also show the predictions we obtain in the context of the Standard Model and in the Standard Model Effective Field Theory including the relevant dimension-six SMEFT operators.
\end{Abstract}
\vfill
\begin{Presented}
$14^\mathrm{th}$ International Workshop on Top Quark Physics\\
(videoconference), 13--17 September, 2021
\end{Presented}
\vfill
\end{titlepage}
\def\thefootnote{\fnsymbol{footnote}}
\setcounter{footnote}{0}

\section{Introduction}

Precision measurements at the LHC can potentially reveal deviations from the Standard Model (SM) expectations, and therefore hint to the existence of new physics phenomena. To exploit future LHC measurements, we have to be able to predict SM processes to our best possible precision. Moreover, a tool by which we can interpret information coming from these measurements is a crucial element to move forward. Assuming new physics is ``hidden" at some scale $\Lambda$ that is currently higher than our reach, the Bottom-Up Standard Model EFT (SMEFT) is one powerful candidate that provides a consistent and model-independent framework where deviations from the SM are encapsulated in higher dimension gauge-invariant operators and can be systematically analysed. 

Even though $tWZ$ is a rare EW process with a cross-section around 100 femtobarns, it does feature a unique unitarity-violating behaviour induced in its sub-amplitudes through the modification of the top quark EW interactions~\cite{Dror:2015nkp,Maltoni:2019aot}. This provides sensitivity to new physics deviations which, despite the complexity of $tWZ$ at NLO accuracy, makes it still an interesting process to study.

\section{$tWZ$ in the SM}
\begin{figure}[h!]
    \centering
    \includegraphics[width=0.7\textwidth]{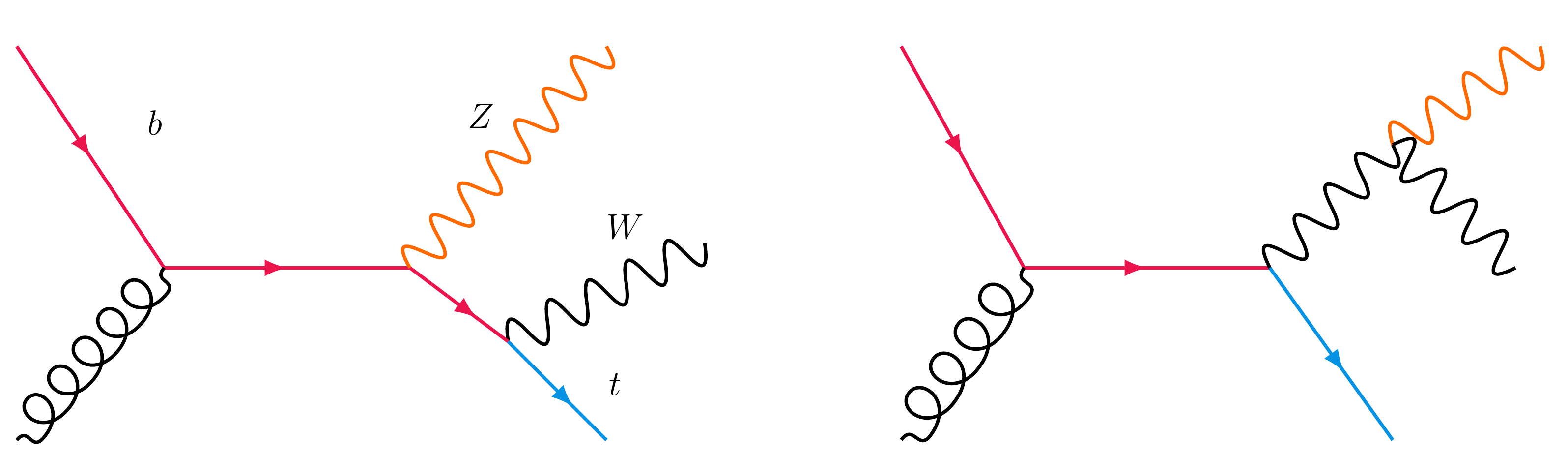}
    \caption{\label{LO}
    {Representative diagrams for $tWZ$ production in the 5FS at leading order.}}
\end{figure}

\begin{figure}[h!]
    \centering
    \includegraphics[width=0.8\textwidth]{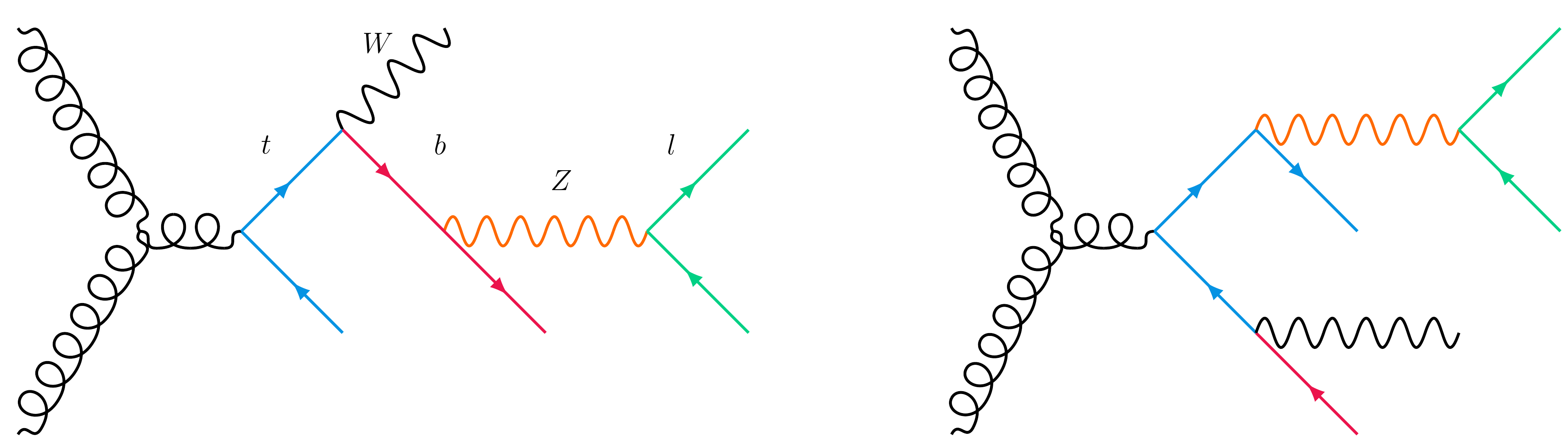}
    \caption{\label{overlap}
    Representative diagrams of the $gg$-initiated production of $t\bar t$ (\emph{left}) and $t\bar{t}Z$ (\emph{right}) processes with a potentially resonant (anti-)top quark leading to the $t\bar \ell \ell Wb$ final state.}
\end{figure}

\begin{figure}[t!]
    \centering
    \includegraphics[trim=0.0cm 5.0cm 0.0cm 0.0cm, clip,width=.48\textwidth]{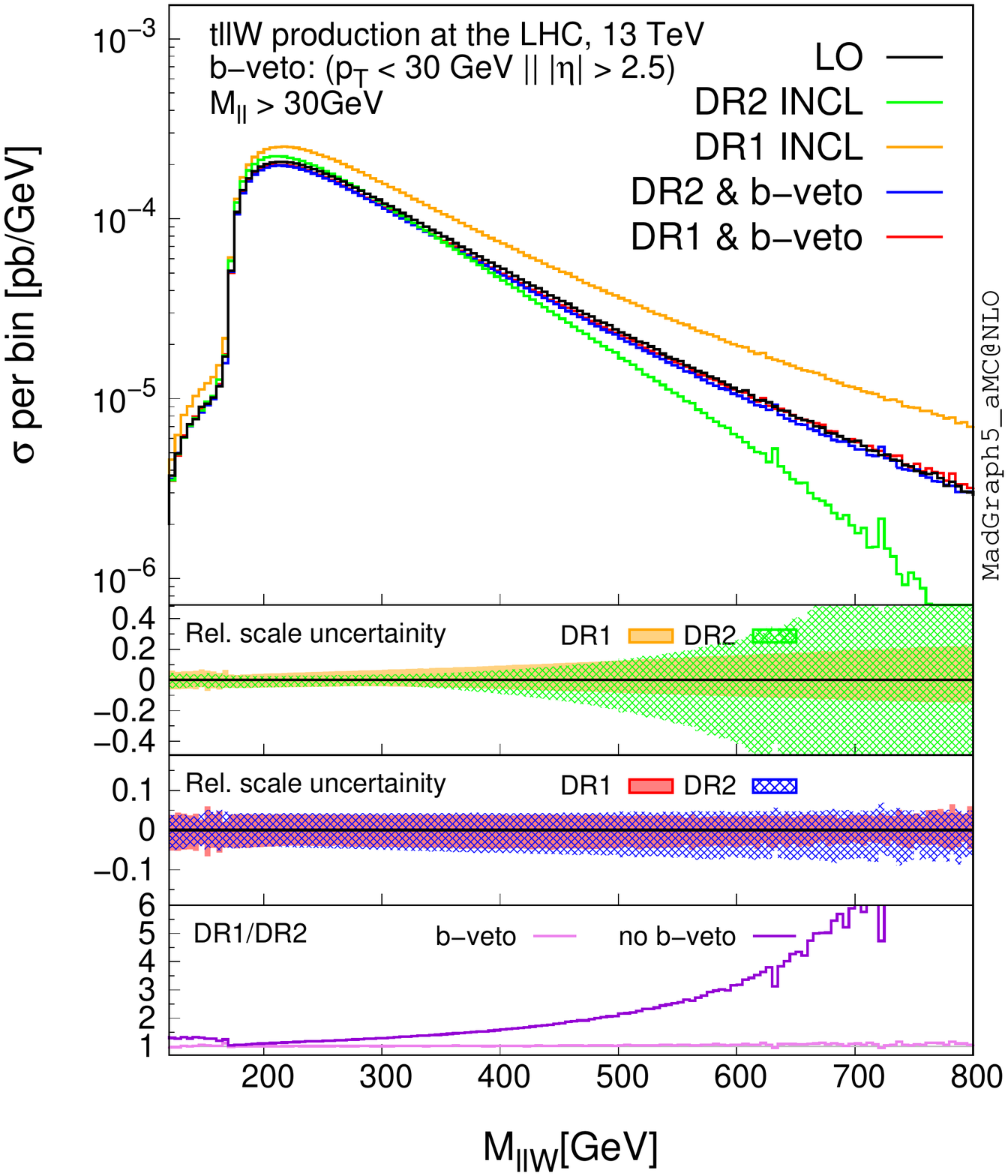}
    \includegraphics[trim=0.0cm 5.0cm 0.0cm 0.0cm, clip,width=.48\textwidth]{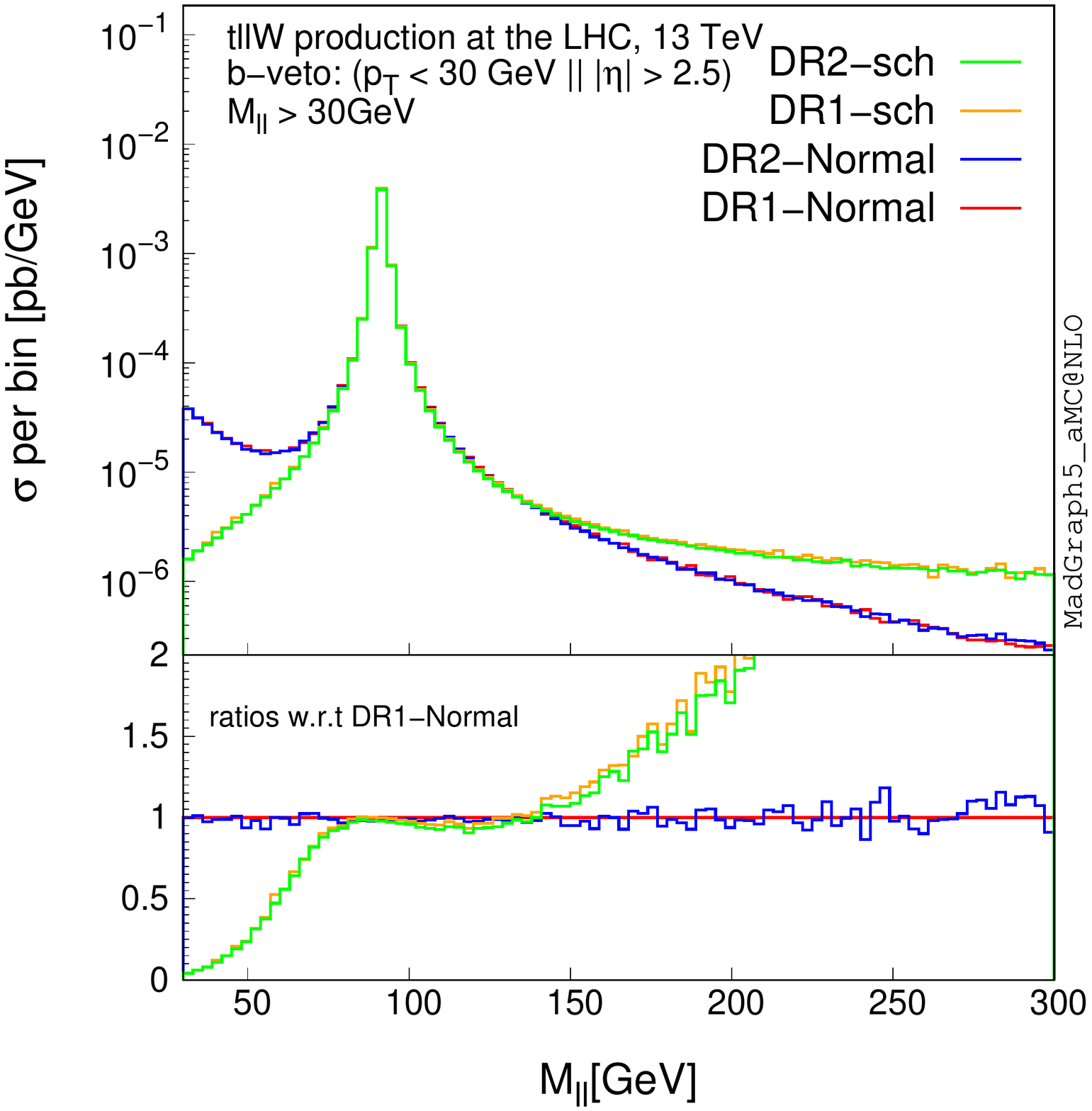}
    \caption{\label{fig:sm_main_text}  The $tWll$ differential cross section in the invariant mass bins of the $Wll$ system (left). The first two insets show the relative scale uncertainties and the third shows the DR1/DR2 ratio. The right panel shows DR1 and DR2 predictions in the invariant mass bins of the lepton pair, $M_{ll}$. The `Normal' case is when we keep all diagrams as opposed to the `sch' where only diagrams with a resonant $Z$ boson are retained. The inset shows the ratios of the distributions w.r.t to the `Normal' DR1 prediction.}
\end{figure}

In the five-flavor scheme (5FS) and at LO, $tWZ$ is identified through the process $gb \to tWZ$ shown in Fig.~\ref{LO}. At NLO and because of the real radiation, a real emission $tWZb$ final state arise from $gg (b \bar b) \to tWZ b$. This real emission final state overlaps with a similar, however resonant, final states induced by an intermediate top quark going on-shell. These resonant final states do not belong to the genuine $tWZ$ production but to $t \bar t$, with $\bar t\to WZ \bar b$, or to $ t\bar t Z$, with $\bar t\to W \bar b$, as displayed in Fig.~\ref{overlap}. For a properly-behaved perturbative expansion and to be able to define the genuine $tWZ$ process at NLO in QCD, such resonant contributions need to be subtracted. 
To this end, we follow the diagram-removal procedures presented in Ref.~\cite{Demartin:2016axk,Frixione:2019fxg}. The amplitude, $\mathcal{M}$, can be decomposed into a non-resonant and a resonant part,
\begin{equation}
    \mathcal{M} = \mathcal{M}_{\rm non-res} + \sum_i \mathcal{M}_{{\rm res}, i},    
\end{equation}
where the index $i$ indicates different resonant processes. The amplitude squared reads
\begin{equation}
    |\mathcal{M}|^2 = \left|\mathcal{M}_{\rm non-res}\right|^2 + 2\Re \left(\mathcal{M}_{\rm non-res} \sum_i \mathcal{M}_{{\rm res}, i}\right) +   \left|\sum_i \mathcal{M}_{{\rm res}, i}\right|^2 .
\end{equation}
In the case of standard DR procedure (DR1), the resonant terms are discarded before squaring the amplitude, i.e.
\begin{equation}\label{eq:DR1}
    |\mathcal{M}|^2_{\rm DR1} \equiv \left|\mathcal{M}_{\rm non-res}\right|^2 .
\end{equation}
In the case of diagram-removal with interference (DR2), the squared resonant matrix elements in the squared amplitude are discarded, but their interference with the non-resonant part is retained,
\begin{equation}\label{eq:DR2}
    |\mathcal{M}|^2_{\rm DR2} \equiv \left|\mathcal{M}_{\rm non-res}\right|^2 + 2\Re \left(\mathcal{M}_{\rm non-res} \sum_i \mathcal{M}_{{\rm res}, i}\right).
\end{equation}
Since the aim is to define $tWZ$ as an independent process, one has to suppress the interference of the non-resonant amplitude with the resonant one, i.e. the second term on the R.H.S of Eq.~\ref{eq:DR2}. We found that vetoing the bottom-tagged jet ($b$-jet) that would stem from the top decay strongly suppresses this term and thus renders DR1 and DR2 predictions in good agreement. The reason why vetoing hard $b$-jets cuts away regions in the phase-space in which $tWZ$ is non-dominant is because $b$-quarks originating from real emission diagrams are instead enhanced in the soft and collinear limits. This is evident in left panel of Fig.~\ref{fig:sm_main_text} which shows the differential cross section predictions in the $Wll$ invariant mass bins for the $tWll$ process at NLO in QCD in the DR1 and DR2 schemes, both at inclusive-level and after applying a veto on $b$-quarks with $|\eta|<2.5$ or $p_T > 30$ GeV.

We proceed by comparing the cross-section of the $tWll$ process in one case where we consider only diagrams with a resonant $Z$ boson to another case where all diagrams are retained. As shown in the right panel of Fig.~\ref{fig:sm_main_text}, we see the two cases in good agreement in the [70, 150] GeV range, and only differ in regions where the cross-section is suppressed. This leads us to simplify our SMEFT study that follows by keeping the $Z$ boson stable. Finally and before moving to the SMEFT study, in Tab.~\ref{tab:sm_results_tab} we show the SM inclusive cross-section for $tWZ$ at LO and at NLO for both the DR1 and the DR2 schemes. 
\begin{table}[ht!]
\centering
\resizebox{0.6\textwidth}{!}{
\renewcommand{\arraystretch}{1.5}
\centering
\begin{tabular}{|c|c|c|c|}
\hline
            & \textbf{LO}                       & \textbf{NLO DR1}                 & \textbf{NLO DR2}                \\ \hline
\textbf{SM} & $103.36(4)^{+12.76\%}_{-12.82\%}$ & $106.70(15)^{+4.97\%}_{-6.28\%}$ & $106.80(9)^{+5.04\%}_{-5.62\%}$ \\ \hline
\end{tabular}
}
\caption{\label{tab:sm_results_tab}The SM contributions [fb] to $tWZ$ production at LO and NLO, including QCD scale uncertainties, for the LHC at $\sqrt{s}=$ 13 TeV.}
\end{table}

\section{$tWZ$ in the SMEFT}
\begin{figure}[h!]
    \centering
    \includegraphics[width=\textwidth]{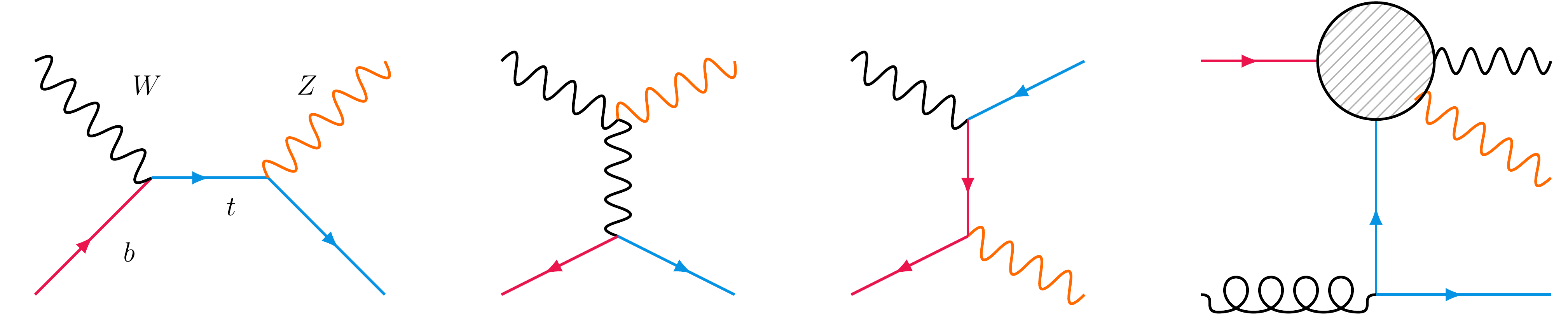}
    \caption{\label{fig:feyn_eft} $b W \to t Z$ sub-amplitudes of the $tWZ$ process. The blob on the far right diagram shows the embedding of the $b W \to t Z$ sub-amplitude into $gb \to tWZ$ production.}
\end{figure}

As previously mentioned, the $b W \to t Z$ sub-amplitudes of the $tWZ$ process shown in Fig.~\ref{fig:feyn_eft} feature a unique sensitivity to unitarity-violating behaviour through modified EW interactions. Therefore, searching for an induced characteristic energy growth can be an interesting path for a new physics search.
Given our sub-amplitudes of interest, the EW top interactions can be modified by the following dimension-six SMEFT operators
\begin{align}
\label{eq:top_ops}
    \Opp{\phi Q}{(3)},\Opp{\phi Q}{(-)},
    \Op{\phi t},\Op{tW},\Op{tZ},\Op{tG}.
\end{align}
The first two \textit{current} operators modify the interactions of two fermions fields with the gauge bosons. The third current operator, $\Op{\phi t}$, is the right handed $t\bar{t}Z$ interaction. The following two \textit{dipole} operators modify the interaction of the top quark with the weak isospin and the weak hypercharge gauge fields. Finally, and even though not included in the $b W \to t Z$ sub-amplitude, $\Op{tG}$ is included in our analysis as it modifies the gluon-top quark interaction.

\begin{figure}[h!]
    \centering
    \includegraphics[trim=0.0cm 5.0cm 0.0cm 0.0cm, clip,width=.45\textwidth]{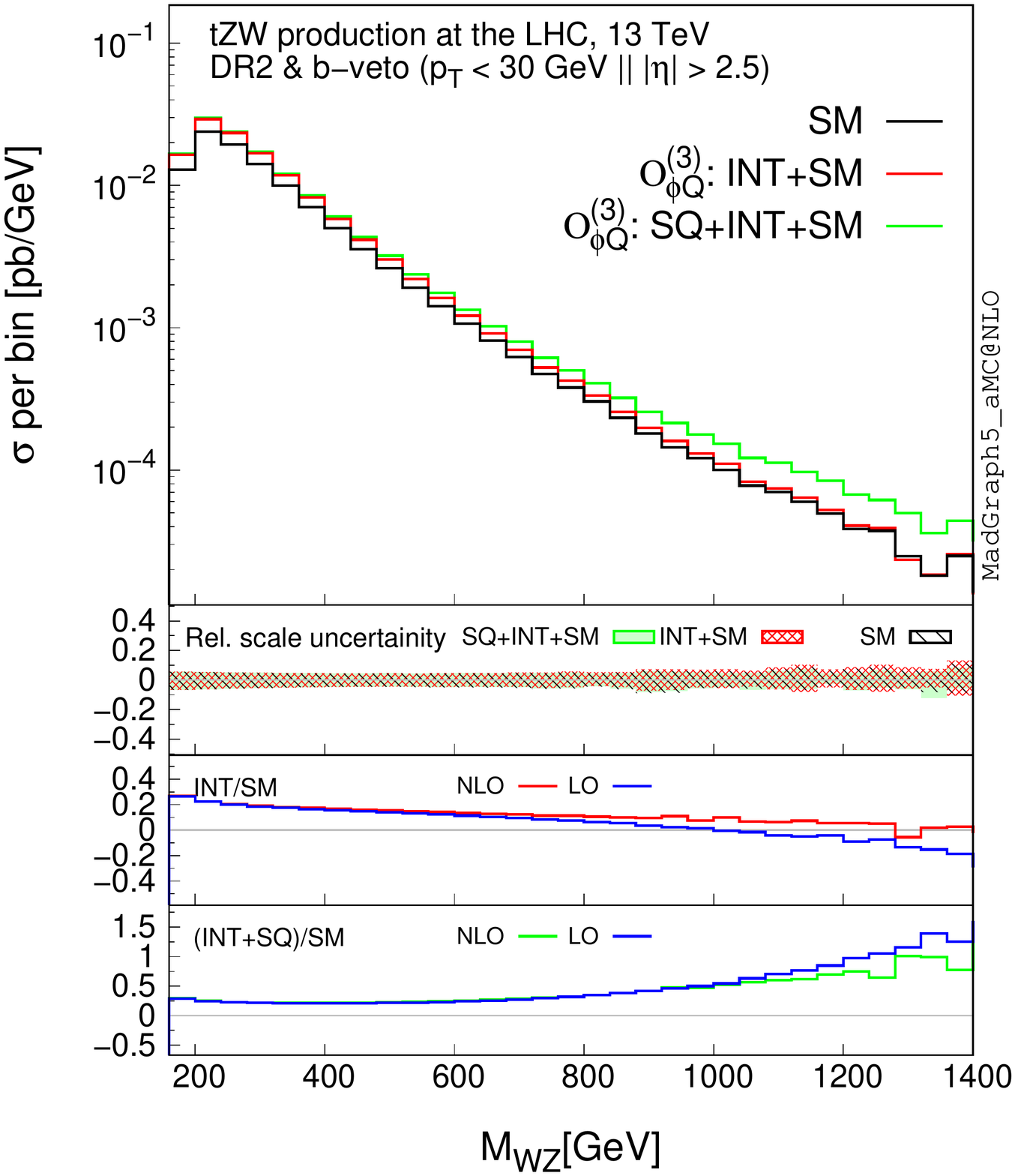}
    \includegraphics[trim=0.0cm 5.0cm 0.0cm 0.0cm, clip,width=.45\textwidth]{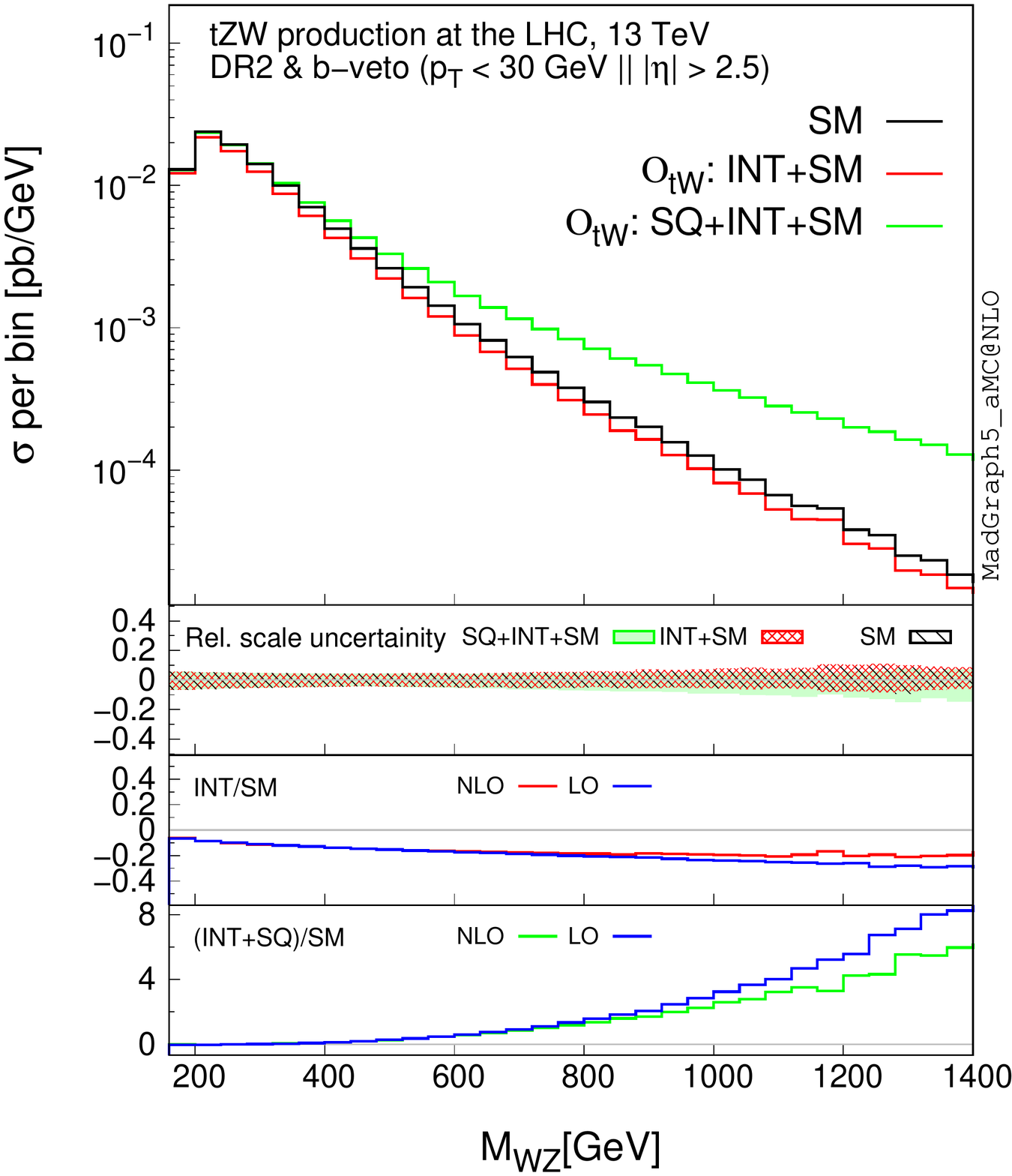}\\
    \includegraphics[trim=0.0cm 5.0cm 0.0cm 0.0cm, clip,width=.45\textwidth]{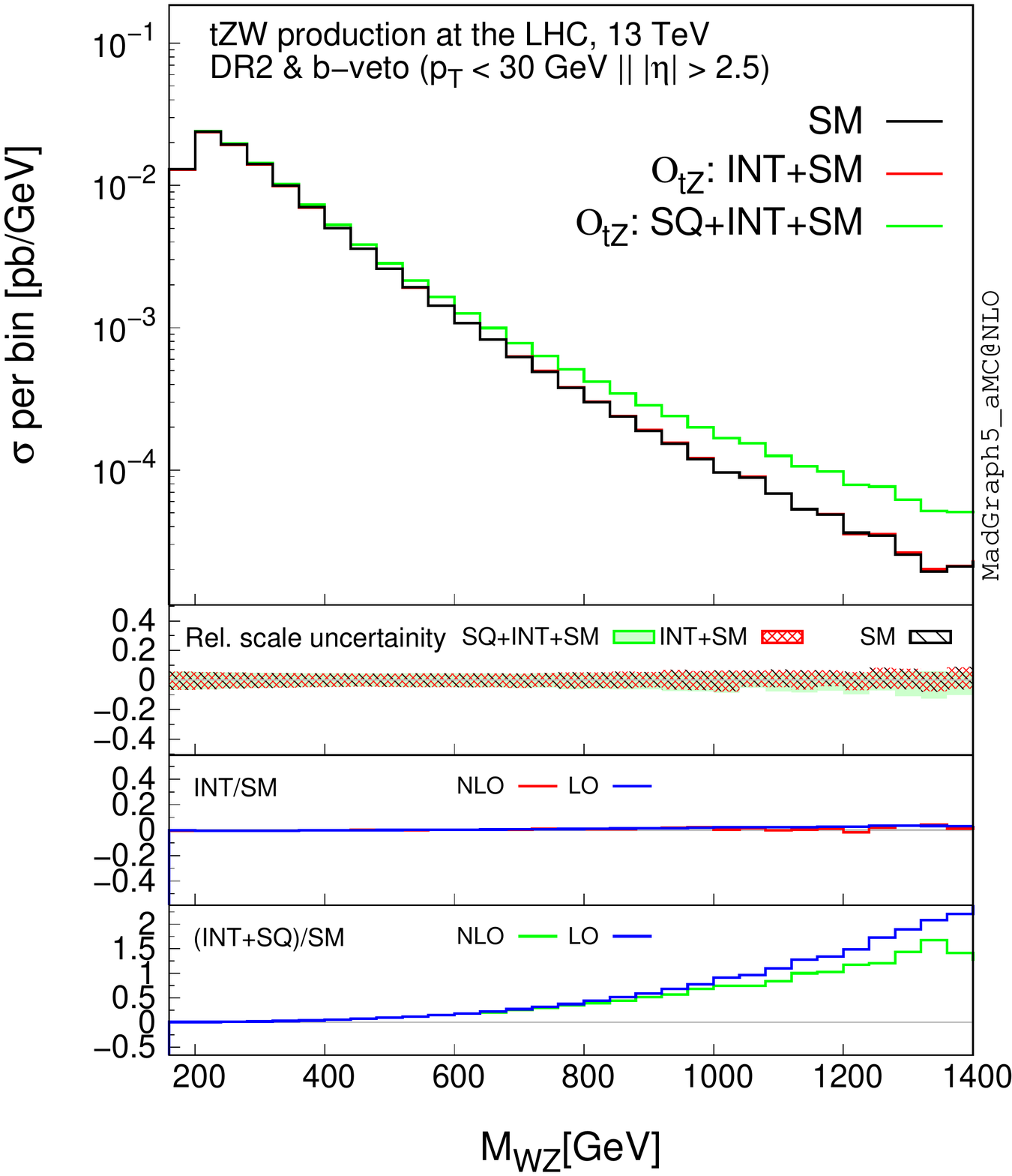}
    \includegraphics[trim=0.0cm 5.0cm 0.0cm 0.0cm, clip,width=.45\textwidth]{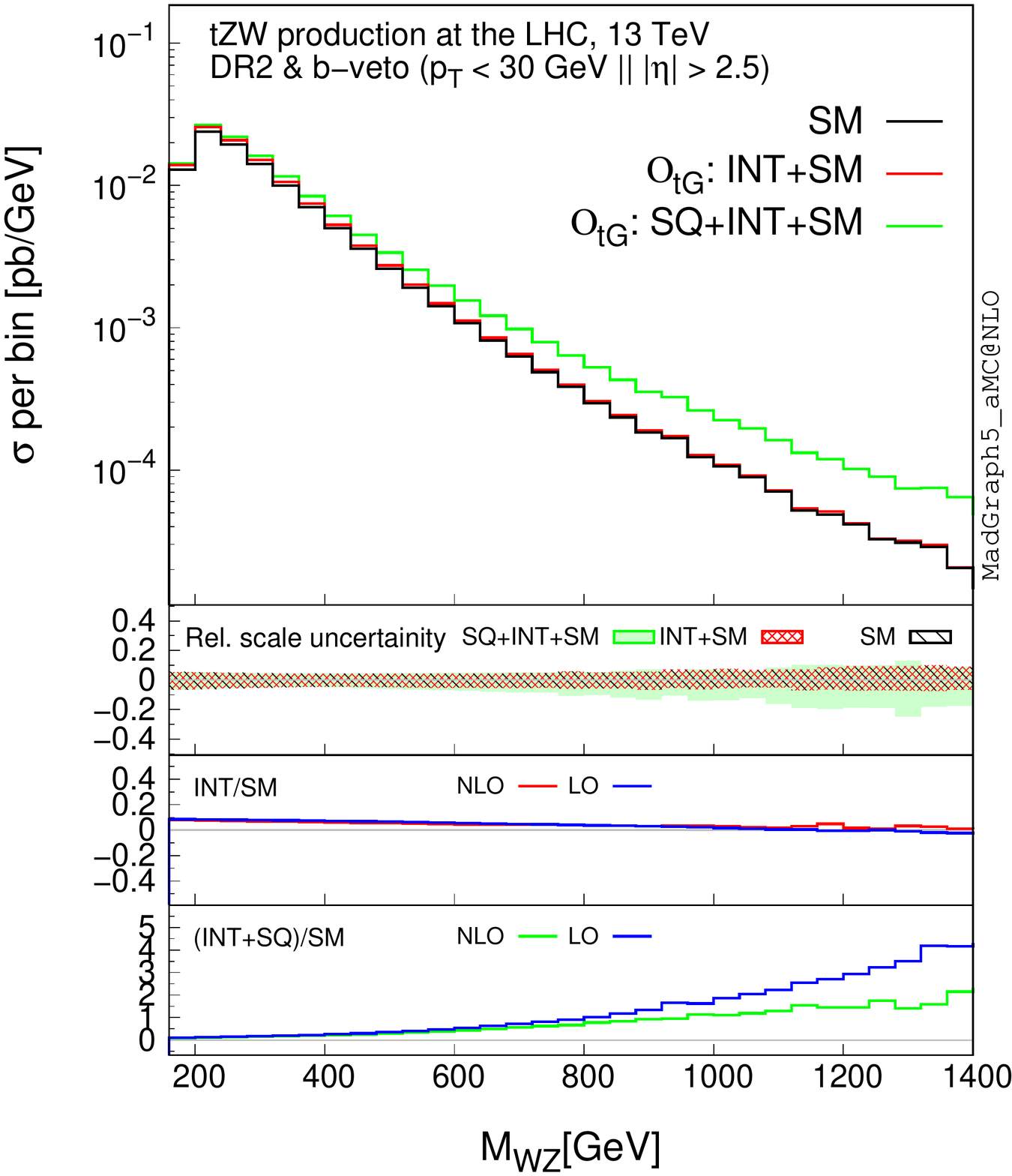}
    \caption{\label{fig:cpq3} The invariant mass of the $WZ$ pair, $M_{WZ}$, at DR2 for the four relevant SMEFT operators when $t\bar{t}Z$ is removed and the $b$-veto is imposed, at FO NLO accuracy. The legends correspond to different computations at different SMEFT contributions. The first inset shows the scale variations in the process. The middle and the last insets show (at LO and at NLO) the interference and the full SMEFT contribution relative to the SM, respectively.}
\end{figure}

It is worth noting that in the high-energy phase-space regions, and since in the SMEFT study we use the on-shell $Z$ approximation as discussed earlier, the contribution of the $t \bar t$ overlap via $\bar t\to WZ \bar b$ decay becomes non-resonant, and thus those $t\bar{t}$ contributions should not be removed. Therefore, in the SMEFT study, only the $t\bar{t}Z$ overlap is of concern. Moving into the SMEFT differential predictions, as with our SM results, the $tWZ$ process is simulated via \texttt{MadGraph5\_aMC@NLO}~\cite{Alwall:2014hca, Frederix:2018nkq} at NLO-QCD accuracy in the SMEFT using \texttt{SMEFTatNLO}~\cite{Degrande:2020evl}. The $b$-veto as discussed in the SM section proves efficient for the SMEFT case in suppressing the resonant contributions to the process and providing stable SMEFT predictions. In Fig.~\ref{fig:cpq3} we show the Fixed Order (FO) DR2 cross-section predictions in the $WZ$ invariant mass bins for four of the six operators presented in Eq.\ref{eq:top_ops}. The reason we omitted $\Opp{\phi Q}{(-)}$ and $\Op{\phi t}$ from our analysis is because they showed a rather insignificant energy growth. 
\section{Summary and conclusions}
I presented a study of the $tWZ$ production process at NLO in QCD where the diagram-removal procedure were used to suppress resonant contributions from LO overlapping processes. We found that vetoing additional hard or central $b$-jet is sufficient to define a region in the phase-space where the genuine $tWZ$ process dominates. 

In the SMEFT study, we considered the EW modifications of the underlying $b W \to t Z$ sub-amplitude through different dimension-six SMEFT operators. The aim of this study is to identify potential energy-growing effects that could be exploited at future collider searches. 

The Fixed Order differential predictions at NLO accuracy for the SM and for the SMEFT including the four relevant operators were presented. Similar to the SM case, we found our SMEFT results to be stable under the diagram-removal procedures. 

\Acknowledgements
H.F is supported by the F.R.S.-FNRS under the “Excellence of Science” EOS be.h project no. 3082081.

\bibliography{main}

\begin{thebibliography}{1}

\bibitem{Dror:2015nkp}
Jeff~Asaf Dror, Marco Farina, Ennio Salvioni, and Javi Serra.
\newblock {Strong tW Scattering at the LHC}.
\newblock {\em JHEP}, 01:071, 2016.

\bibitem{Maltoni:2019aot}
Fabio Maltoni, Luca Mantani, and Ken Mimasu.
\newblock {Top-quark electroweak interactions at high energy}.
\newblock {\em JHEP}, 10:004, 2019.

\bibitem{Demartin:2016axk}
Federico Demartin, Benedikt Maier, Fabio Maltoni, Kentarou Mawatari, and Marco
  Zaro.
\newblock {tWH associated production at the LHC}.
\newblock {\em Eur. Phys. J. C}, 77(1):34, 2017.

\bibitem{Frixione:2019fxg}
Stefano Frixione, Benjamin Fuks, Valentin Hirschi, Kentarou Mawatari, Hua-Sheng
  Shao, P.~A. Sunder, and Marco Zaro.
\newblock {Automated simulations beyond the Standard Model: supersymmetry}.
\newblock {\em JHEP}, 12:008, 2019.

\bibitem{Alwall:2014hca}
J.~Alwall, R.~Frederix, S.~Frixione, V.~Hirschi, F.~Maltoni, O.~Mattelaer,
  H.~S. Shao, T.~Stelzer, P.~Torrielli, and M.~Zaro.
\newblock {The automated computation of tree-level and next-to-leading order
  differential cross sections, and their matching to parton shower
  simulations}.
\newblock {\em JHEP}, 07:079, 2014.

\bibitem{Frederix:2018nkq}
R.~Frederix, S.~Frixione, V.~Hirschi, D.~Pagani, H.~S. Shao, and M.~Zaro.
\newblock {The automation of next-to-leading order electroweak calculations}.
\newblock {\em JHEP}, 07:185, 2018.
\newblock [Erratum: JHEP 11, 085 (2021)].

\bibitem{Degrande:2020evl}
C\'eline Degrande, Gauthier Durieux, Fabio Maltoni, Ken Mimasu, Eleni
  Vryonidou, and Cen Zhang.
\newblock {Automated one-loop computations in the standard model effective
  field theory}.
\newblock {\em Phys. Rev. D}, 103(9):096024, 2021.

\end{thebibliography}
\bibliographystyle{unsrt}
 
\end{document}